\documentclass[aoas,MSNbibl,nameyear,rotating,dvips]{arximspdf}
\usepackage{dcolumn}
\usepackage{graphicx}

\doi{10.1214/14-AOAS780} 
\volume{8}
\issue{4}
\pubyear{2014}
\firstpage{2068}
\lastpage{2095}
\docsubty{FLA}

\makeatletter
\newcolumntype{d}[1]{D{.}{.}{#1}}
\newcommand{\rrvert}{\vert}
\newcommand{\rrVert}{\Vert}
\newcommand{\llvert}{\vert}
\newcommand{\llVert}{\Vert}
\renewcommand{\mid}{|}
\newcommand{\tr}{\operatorname{tr}}
\newcommand{\me}{\mathrm{e}}
\newcommand{\reals}{\mathsf{R}}
\newcommand{\SO}{\mathsf{SO}}
\newcommand{\normal}{\mathsf{N}}
\newcommand{\Gam}{\mathsf{Ga}}
\newcommand{\IGam}{\mathsf{IGa}}
\makeatother

\begin{document}
\begin{frontmatter}

\title{Bayesian protein structure alignment\thanksref{T1}}
\runtitle{Bayesian protein structure alignment}

\begin{aug}
\author[A]{\fnms{Abel}~\snm{Rodriguez}\corref{}\ead[label=e1]{abel@soe.ucsc.edu}}
\and
\author[B]{\fnms{Scott C.}~\snm{Schmidler}}
\runauthor{A. Rodriguez and S. C. Schmidler}
\affiliation{University of California, Santa Cruz and Duke University}
\address[A]{Department of Applied Mathematics\\
\quad and Statistics\\
University of California, Santa Cruz\\
1156 High Street\\
Mailstop SOE2\\
Santa Cruz, California 95064\\
USA\\
\printead{e1}} 
\address[B]{Department of Statistical Science\\
Duke University\\
Box 90251\\
Durham, North Carolina 27708\\
USA}
\end{aug}
\thankstext{T1}{Supported in part by NSF Grant DMS-06-05141 (SCS) and
NIH/NIGMS R01GM090201-01.}

\received{\smonth{3} \syear{2014}}
\revised{\smonth{7} \syear{2014}}

%
\begin{abstract}
The analysis of the three-dimensional structure of proteins is an
important topic in molecular biochemistry. Structure plays a
critical role in defining the function of proteins and is more
strongly conserved than amino acid sequence over evolutionary
timescales. A key challenge is the identification and evaluation of
structural similarity between proteins; such analysis can aid in
understanding the role of newly discovered proteins and help
elucidate evolutionary relationships between organisms.
Computational biologists have developed many clever algorithmic
techniques for comparing protein structures, however, all are based
on heuristic optimization criteria, making statistical interpretation
somewhat difficult. Here we present a fully probabilistic framework
for pairwise structural alignment of proteins. Our approach has
several advantages, including the ability to capture alignment
uncertainty and to estimate key ``gap'' parameters which critically
affect the quality of the alignment. We show that several existing
alignment methods arise as \textit{maximum} {a posteriori} estimates
under specific choices of prior distributions and error models. Our
probabilistic framework is also easily extended to incorporate
additional information, which we demonstrate by including primary
sequence information to generate simultaneous sequence--structure
alignments that can resolve ambiguities obtained using structure
alone. This combined model also provides a natural approach for the
difficult task of estimating evolutionary distance based on
structural alignments. The model is illustrated by comparison with
well-established methods on several challenging protein alignment
examples.
\end{abstract}

%
\end{frontmatter}

\section{Introduction}\label{sec1}
Protein alignment is among the most powerful and\break widely used tools
available for inferring homology and function of gene products, as
well as determining evolutionary relationships between organisms. In
particular, protein sequence alignment uses information about the
identity of amino acids to establish regions of similarity, and has a
long history of providing valuable insights. For example, the
alignment of a putative human colon cancer gene with a yeast mismatch
repair gene played a crucial rule in its identification and
characterization [\citet{Br94,Pa94,ZhLiLa98}].

Sequence alignment is most useful for shorter evolutionary distances,
when amino acid composition has not drifted dramatically from a common
ancestor. However, when comparing proteins that are distantly
related, sequence conservation may be too dilute to establish
meaningful relationships. Because a protein's function is largely
determined by its three-dimensional structure, and significant
sequence mutation can occur while maintaining this structure, it is
widely recognized that structural similarity is conserved over much
longer evolutionary timescales than sequence similarity. In addition,
sequence alignment cannot detect convergent evolution, when proteins
with similar 3D structure and carrying out similar functions have
evolved from unrelated genes.

Aligning 3D structures requires choosing which amino acids to match as
in sequence alignment, but has the added complexity of handling
coordinate frames arising from arbitrary rotation and translation.
Early work in structural alignment [\citet{RaRo73}, \citeauthor{RoAr75} (\citeyear{RoAr75,RoAr76})]
developed techniques that iterate between a rigid body registration
and an alignment step, and \citet{SaCoPaAv86} introduced the use of
dynamic programming [applied to sequence alignment by \citet{NeWu70}]
as an efficient way to construct the alignment given a registration.
Similar methods have been adopted by many authors
[\citet{Co97,GeLe98,Wu1998b}]. Most work uses a penalized root mean
squared deviation (RMSD) between corresponding backbone
$\alpha$-carbon (C$_\alpha$) atoms to measure quality of the
alignments, but several other measures have been proposed, including
soap-bubble surface metrics [\citet{FaCo96}], differential geometry
[\citet{KoNiTe03}], and heuristic rules like the SSAP method of
\citet{TaOr89}.

An alternative to iterative methods is the use of distance geometry
to avoid the registration problem, thus representing each protein by a
pairwise distance matrix between all $C_\alpha$ atoms. The popular
DALI [\citet{dali93}] method is an example of this approach. Other
techniques are specially tailored for the large-scale computational
demands of rapid searching of large protein databases, sometimes
employing highly redundant representations of the data; these include
geometric hashing [\citet{AlGiMiMyLi90,FiWoLiNu94,WaLaTh96}], graph
algorithms [\citet{Ta02}] and clustering methods like VAST
[\citet{vast96}]. Finally, some authors combine these ideas with additional
heuristics to produce faster or more accurate algorithms, including CE
[\citet{ShBo98}] and PROSUP [\citet{LaKoSiDo00}]. Detailed reviews on
pairwise structural alignment methods can be found in \citet{BrOrTa96},
\citet{EiJoTa00} and \citet{LeLe00}.

The profusion of methods shows the difficulties involved in performing
structural alignments: in defining how to measure alignment quality
and in computing ``best'' alignments efficiently. It has been well
documented in the literature that different algorithms can produce
alignments sharing very few amino acid pairings, and are sensitive to
both the initial alignment and the specific choice of algorithm
parameters [\citet{Go96,ZuSi96,GeLe98}]. Additional complications arise
when trying to determine the significance of the resulting
alignments. Although substantial effort has been devoted to this\vadjust{\goodbreak}
point and important progress made [\citet{LiPe85,MiGo95,LeGe98,GeLe98}],
the solutions remain based on heuristics and upper bounds that are
difficult to interpret. Finally, all the methods
described above approach the structural alignment as an optimization
problem, finding a single best alignment. However, structural
comparisons are subject to substantial uncertainties arising from
evolutionary divergence, population variability, experimental
measurement error and protein conformational variability, not to
mention sensitivity to parameters of comparison metrics and optimization
algorithms. To address these sources of variability, approaches
based on explicit statistical modeling are desirable, and the results
of structural comparisons require careful analysis to understand the
impact of uncertainty.

In this paper, we develop a Bayesian statistical approach to
pairwise protein structure alignment, combining techniques from
statistical shape analysis [\citet
{Dryden1998,Small1996,Kendall1999}] and
Bayesian sequence alignment [\citet{ZhLiLa98,WeLiLa98,LiLa99}]. This
represents one aspect of a general Bayesian framework developed here
and elsewhere [\citeauthor{Schmidler2003} (\citeyear{Schmidler2003,Schmidler2004})], and subsequently
extended by \citeauthor{Schmidler2006} (\citeyear{Schmidler2006,Schmidler2005}),
\citet{Wang2006}.
\citet{GrMa06} and \citet{DrHiMe06} independently developed
related approaches for
hierarchical Bayesian alignment of protein active sites rather than
whole proteins, and for small molecules, respectively. However, our
approach differs in a number of important points: we introduce
hierarchical priors on the space of alignments that are equivalent to
the standard affine gap penalty of classical alignment approaches, but
allow us to estimate the parameters controlling the complexity of the
alignment. We also introduce an efficient computational approach that
allows rapid computation and sampling, which both enables
identification of
alternative alignments and provides direct measures of alignment
uncertainty. A
significant advantage of our formulation is the unification of many
existing alternative methods for structural alignment, which can be
seen as special cases of MAP alignment under different specific
choices of error models or alignment priors. This provides valuable
insight into the relationships and properties of existing algorithms.

Another powerful advantage of a fully probabilistic framework is the
ability to incorporate disparate sources of information in a natural
and coherent fashion. Using our Bayesian structural alignment model
as a platform, we also develop a fully probabilistic approach for \textit{simultaneous sequence-and-structure} alignment, which combines
information from both primary sequence \textit{and} 3D structure. In the
presence of unambiguity in geometric matching for highly-divergent
proteins or low-resolution structural data, amino acid identities or
preferred substitutions can significantly alleviate the remaining
uncertainty.
We demonstrate this approach on some difficult structural alignment
problems from the literature. Finally, we show that our simultaneous
alignment approach provides a natural method for estimating
evolutionary distances directly from structure comparison, a
notoriously difficult task.

\section{Proteins and their structure}\label{sec2}

\begin{figure}[t]

\includegraphics{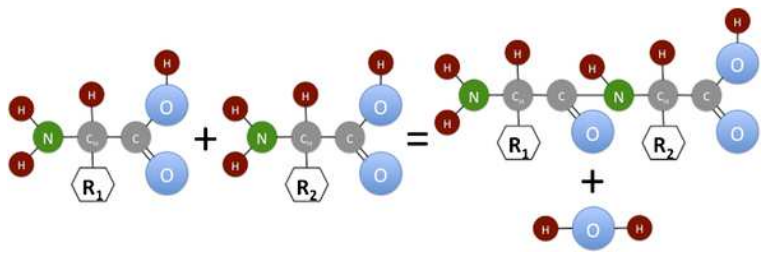}

\caption{The chemical structure of proteins, showing the combination
of two amino acids to form a peptide bond. Repeated applications of
this process form a linear chain to make a protein. The identities
of the R-groups determine the protein sequence and thus its
properties, including 3D structure and biochemical function.}\label{fiaminoacidchem}
\end{figure}

\begin{figure}[b]

\includegraphics{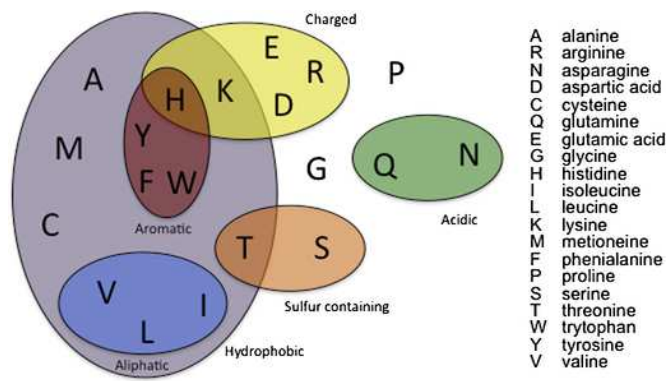}

\caption{The twenty amino acids encoded by the universal genetic code,
and their chemical properties.}\label{fiaminoacidprop}
\end{figure}

Proteins are the most diverse macromolecules in organisms, playing a
wide range of roles: as enzymes, molecular receptors, antibodies,
hormones, structural proteins, and molecular transporters. Proteins are
linear polymers, molecular chains created by stringing together amino
acids using peptide bonds to form a polypeptide. The constituent amino
acids are themselves small molecules characterized by a central carbon atom
(C$_{\alpha}$) to which additional chemical groups are attached,
including a carboxyl group (COOH), an amino group (NH$_2$) and an
organic side chain (see Figure~\ref{fiaminoacidchem}). There are 20
distinct naturally occurring types of
side chains, ranging from very simple (G) to relatively complex (F),
giving the 20 naturally-occurring amino acids their identities.
During the process of peptide-bond formation, a water molecule is shed
and, as a
result, amino acids occurring within a protein chain are often referred
to as ``residues.'' Because residues are not symmetric, the chain is
directional, with the beginning end having a free amino group known as the
amino- (or N-) terminus and the end having a free carboxyl group known
as the
carboxy- (or C-) terminus. The sequences of amino acids making up
proteins are encoded in DNA by the universal genetic code; Figure~\ref{fiaminoacidprop} shows a simple classification of these
amino acids including some of their chemical properties. It is the
combinatorics of combining these properties in different numbers and
orderings that gives rise to the diversity of protein structures and functions.
%
%

%

The linear sequence of amino acid identities makes up the \textit{primary} structure of a protein and, like DNA, can be encoded using
strings of letters. Primary protein sequences can be aligned to
identify evolutionarily related or otherwise similar regions, using
algorithms for string
comparison. This requires an amino acid distance metric, often
summarized in the form of substitution matrices such as PAM
[\citet{Dayhoff68}] or BLOSUM [\citet{Henikoff92}]. Sequence alignments
can provide
important insights into the function of proteins and the evolution of organisms.

The diverse chemical properties of amino acids lead proteins
to ``fold'' reproducibly into complicated, sequence-specific bundles.
This three-\break dimensional structure enables a protein to
perform its functions (such as specific binding of a target). Within
this fold are often smaller, recognizable structural ``motifs''
occurring across many proteins, known as secondary structures. These are
regularly repeating local structural patterns, with the most famous
being $\alpha$-helices (successive backbone atoms following a right-handed
helical path though space) and $\beta$-sheets
(extended stretches of backbone connected laterally to form sheets).
Because these secondary structure elements are local, many regions
of different secondary structure can be present in the same protein
molecule. In contrast, the \textit{tertiary} structure of a protein
refers the overall 3D shape, including the relative locations of
secondary structures in space. In this paper we are concerned with
the problem of alignment between this 3D structure of two proteins, as
3D structures tend to be much more highly conserved across evolution
than the
sequence itself. This 3D shape is well summarized by the positions of
the C$_{\alpha}$ carbons, giving a path through space known as the
backbone of the protein.

Protein structural data arises most frequently from the
experimental methods of X-ray crystallography and NMR spectroscopy.
Although these experimental techniques differ greatly, the end result
of each is a set of 3D coordinates for the protein atoms in an
arbitrary coordinate system.
These coordinates, which are publicly available at the Protein Data Bank repository
(\surl{http://www.pdb.org/}) along with the primary sequence, are the
data we use in developing our models.

\section{Bayesian protein structure alignment}\label{sec3}

Let $X_{n \times3}$ and $Y_{m \times3}$ be coordinate matrices for
two proteins,
with rows $x_i$ ($y_i$) containing coordinates of the C$_\alpha$ of the
$i$th amino acid. An alignment between $X$ and $Y$ is a
$n \times m$ match matrix $M = (m_{ij})$ such that $m_{ij} = 1$ if
residues $X_i$
and $Y_j$ are matched, and 0 otherwise. Each position in $X$ can
be matched to at most one position in $Y$, so each row and column of $M$
contains at most one nonzero entry, thus, $M$ is the adjacency matrix
for a
matching (a subset of edges, no two sharing an endpoint) on a
complete bipartite graph between sets $X$ and $Y$. In the sequel we
denote by $X_M$ and $Y_M$ the
$\llvert M\rrvert=\sum_{ij}m_{ij}$ nonzero rows of $M'X$ and
$MY$ giving the
coordinates of the matched positions, and by $X_{\bar{M}}$ and
$Y_{\bar{M}}$ the rows of $X$ and $Y$ not included in $X_M$ and
$Y_M$ giving coordinates of the unmatched position.

We adopt a Bayesian approach to structure alignment which defines a
prior distribution on alignments $P(M)$ and, given a probability model
for the coordinates matrices $X$ and $Y$ conditional on $M$, obtains
the posterior distribution:
\[
P(M \mid X, Y) = \frac{P(X,Y \mid M)P(M)}{\sum_M P(X,Y \mid M)P(M)},
\]
where the marginal likelihood $P(X,Y)=\sum_M P(X,Y \mid M)P(M)$
involves a sum over all possible alignments. Although the number of
matchings is exponential in $n$ and $m$, inferences may be obtained by
Monte Carlo sampling from the posterior $P(M \mid X,Y)$ to approximate
posterior summaries such as the posterior mode,
%
%
\begin{equation}
\hat{M} = \arg\max_M P(M \mid X,Y), \label{EqnMAP}
\end{equation}
or the \textit{marginal alignment matrix}, $(p_{ij})$, giving the marginal
posterior
probability $p_{ij}=\sum_M m_{ij}P(M\mid X,Y)$ of matching position
$X_i$ with $Y_i$ summing over all
possible alignments. MAP estimates have well-known
drawbacks: they ignore the variability in the posterior, including that
arising from
uncertainty in hyperparameters or potential multimodality. However,
they are
simple to obtain and provide a convenient ``representative'' alignment
in which each residue matches at most one other. The
marginal alignment matrix is easily obtained by sampling alignments
from the posterior distribution, but is somewhat harder to
visualize. In Section~\ref{seexamples} we use heatmaps for this
purpose, but it is also possible to generate a point estimator by
maximizing an appropriate utility function, as an alternative to the
MAP alignment.

\subsection{Likelihood}\label{sec3.1}\label{selik}

Given a matching matrix $M$, we factor the joint likelihood of the
observed structures $X$ and $Y$ into (dependent) matched and
(independent) unmatched positions:
%
%
\begin{equation}
P(X,Y \mid M) = P(Y_M \mid X_M) P(Y_{\bar{M}})
P(X). \label{EqnFactor}
\end{equation}
This arises naturally, for example, by viewing the aligned positions as
homologous (having a
common evolutionary ancestor) and the unmatched positions as random
insertions and deletions occurring independently in each protein after
divergence. Note that while assuming $Y_M \bot
Y_{\bar{M}} \mid M$ ignores the physical constraints of
neighboring bonds, it simplifies the calculations in important ways
described below.

We adopt a probabilistic model for matched regions of the proteins
which assumes that deviations are independent and normally
distributed,
%
%
\begin{equation}
Y_M = X_M + \varepsilon, \qquad\varepsilon\sim\normal
\bigl(\mathbf{0},\sigma^2I\bigr), \label{EqnShapeModel}
\end{equation}
and, to complete the model, we assume that the rows in $Y_{\bar{M}}$ follow
common distribution $f$. (This normality assumption is discussed
with possible relaxations in Section~\ref{seinter}.)

However, both $X$ and $Y$ are observed only up to arbitrary coordinate
frames. That is, we observe $[X]$ and $[Y]$, where $[X]$ denotes the
\textit{size-and-shape} of $X$, formally defined
[\citet{Dryden1998,Kendall1999}] as an equivalence class of invariant
matrices under the group of Euclidean transformations:
\[
[X] = \bigl\{ XR + \mu\dvtx  R \in\SO(3), \mu\in\reals^3 \bigr\}.
\]
Here $\SO(3)$ is the special orthogonal group of $3\times3$ rotation
matrices. [Alignment using a somewhat more general class of nonrigid
transformations is considered by \citet{Schmidler2005}.]
We therefore define the likelihood by
%
%
\begin{eqnarray}\label{EqnStructJoint}
\qquad P(X,Y \mid M)
&=&  \bigl(2\pi\sigma^2\bigr)^{-{3\llvert M\rrvert}/{2}}
\nonumber\\[-8pt]\\[-8pt]\nonumber
&&{}\times \exp{-\frac{1}{2\sigma^2} \bigl\llVert Y_M - \bigl(X_M
\hat{R}_M + \mathbf{1}\hat{\mu}_M'\bigr)
\bigr\rrVert_F^2} P(X) \prod
_{y_i \in Y_{\bar{M}}} f(y_i \mid\lambda),
\end{eqnarray}
where $\llVert X\rrVert_F= \tr(X'X)^{{1}/{2}}$ is the
Frobenius norm,
$f( \cdot;\lambda)$ is a one-parameter density for inserted/deleted
positions, and $P(X)$ is a\vspace*{1pt} probability distribution describing the
shape of
the reference protein $X$. Here $(\hat{R}_M,\hat{\mu}_M)$ are the
optimal least-squares rotation and translation placing $X$ and $Y$ on a\vspace*{1pt} common
coordinate system, given by $\hat{R}_M=
U_MV_M^T$ and $\hat{\mu}_M = \bar{Y}_M - \bar{X}_M\hat{R}_M$,
where $V_M,U_M \in
\SO(3)$ are obtained from the singular value decomposition $Y_M^TC_MX_M =
U_MD_MV_M^T$ for centering matrix $C_M=I -\frac{1}{\llvert M\rrvert}11^T$.

The appearance of $(\hat{\mu}_M,\hat{R}_M)$ in likelihood
(\ref{EqnStructJoint}) may be interpreted in two different
ways. First, (\ref{EqnStructJoint}) may be viewed as a \textit{profile} likelihood for $M$, maximizing over nuisance parameters
$(\mu,R)$ corresponding to the unknown translation and
rotation conditional on $M$, under the model
\[
Y_M = (X_M + \varepsilon)R + \mu, \qquad\varepsilon
\sim\normal\bigl(\mathbf{0},\sigma^2I\bigr).
\]
A fully Bayesian approach would instead assign prior
distributions to these nuisance parameters and integrate them out.
\citet{GrMa06}, \citet{DrHiMe06} and \citet{Wang2006}
adopt this integration approach, and \citeauthor{Schmidler2006} (\citeyear{Schmidler2006,Schmidler2005})
considers both (maximization and integration)
approaches for handling the unknown registration parameters.
However, in our experience the uncertainty on $(\mu,R)$
given $M$ is minimal for most structural alignments, with the
posterior heavily concentrated about the mode, making the two
approaches perform nearly identically. \citet{kenobi2012bayesian}
report similar
findings and discuss this issue in detail.

We may also interpret (\ref{EqnStructJoint}) as a sampling
density defined directly on (a local tangent space approximation to)
the underlying shape space of the configurations, replacing $3\llvert
M\rrvert$
in the normalizing constant with $3\llvert M\rrvert - 6$, the dimension
of the
shape manifold. The exponent $\llVert Y_M - (X_M\hat{R}_M +
\mathbf{1}\hat{\mu}'_M)\rrVert_F^2=d_P^2(X,Y)$ is known as the
(squared)
\textit{partial Procrustes distance}, and serves as the Riemannian metric
on this (size-and) shape space [\citet{Dryden1998,Kendall1999}].
This metric effectively \textit{defines} a one-to-one correspondence
between matchings $M$ and Euclidean transformations $(\mu,R)$, enabling
inference to be performed directly in the space of matchings.
Under this interpretation, our approach is
fully Bayesian, but the likelihood is approximated by evaluating the
density in the tangent space.

In what follows, we take $f(\cdot\mid\lambda) = \lambda= 1/\llvert
\Omega\rrvert$
uniform over a bounded region~$\Omega$; then $\lambda$ can be
interpreted as a lower bound for the gap penalty as discussed in
Section~\ref{seinter}. Note that the factorization
(\ref{EqnFactor}) implies that the marginal distribution $P(X)$
cancels in the posterior
distribution, and may be left unspecified so long as it is assumed to be
functionally independent of parameters $M$ and $\sigma^2$. This is
similar to
a proportional hazards model where the baseline risk is left
unspecified to obtain a semi-parametric survival model.
In addition, the isotropic error model ensures the model is symmetric
in $X$ and $Y$ if we take
%
%
\[
P(X) = \prod_{x_i \in X}
f(x_i \mid\lambda).
\]
%

\subsection{Prior on the alignment matrix}\label{sec3.2}
Prior distributions on matchings $P(M)$ may be specified in a variety
of ways; here we adopt a \textit{gap-penalty} formulation
familiar in the sequence and structure alignment literature, where
unmatched stretches of amino acids are penalized by the affine function:
\[
u(M;g,h) = g s(M) + h \sum_{i=1}^{s(M)}
l_i(M)
\]
with gap-opening penalty $g$ and gap-extension penalty $h$, where
$s(M)$ is the number of gaps in alignment $M$ and $l_i(M)$ is the
length of the $i$th gap. Exponentiating and normalizing this
function provides a prior on $M$ [\citet{LiLa99}], essentially a Markov
chain parametrized as a ``Boltzmann chain'' Gibbs random field
[\citet{Saul1995,Schmidler2005f}]:
%
%
\begin{equation}
P(M \mid g,h) = Z(g,h) \me^{-u(M;g,h)} \label{eqpriormatrix}
\end{equation}
with normalizing constant $Z$. This prior encourages grouping of
matches together along the protein backbone. It allows for explicit
control over the number of gaps, compared to, for example, the prior of
\citet{GrMa06} which controls only the expected total length.

Under the affine-gap-penalty prior, sampling may be done efficiently
using stochastic recursions analogous to those of standard sequence
alignment algorithms [\citet{LiLa99}], along with additional Monte Carlo
steps, as described below. Note that this prior requires the alignment
to preserve the sequential order along the polypeptide backbone,
requiring topological equivalence of the two proteins. More general
priors applicable for comparing proteins of potentially different
topologies (convergent evolution) are easily accommodated with the
introduction of additional Monte Carlo steps, but will be described
elsewhere. Although the prior allows for simultaneous gaps on both
proteins, for identifiability purposes we do not allow gaps in $X$ to
be followed by gaps in $Y$ [see \citet{WeLiLa98} for details].

\subsection{Hyperpriors}\label{sec3.3}
In standard sequence and structure alignment algorithms, the gap
parameters $g$ and $h$ are assigned fixed values. However, they have
a critical effect on the resulting alignment, with large
opening gap penalties $g$ tending to produce alignments with few gaps and
vice versa. In the context of sequence alignment, \citet{LiLa99} treat
$(g,h)$ as nuisance parameters and assign hyperpriors, integrating them
out to obtain a marginal posterior distribution over alignments.
We similarly assign $g$ and $h$ Gamma hyperpriors,
%
%
\begin{equation}
g \sim\Gam(a_g, b_g), \qquad h \sim
\Gam(a_h, b_h), \label{eqpriorgh}
\end{equation}
with hyperparameters $(a_g,b_g,a_h,b_h)$ chosen to be diffuse (but
proper). An
alternative is to utilize manually-obtained reference alignments [e.g., BAliBASE, see \citet{ThPlPo99} and \citet{ThKoRiPo05}] to
obtain informative
priors for $g$ and $h$. The\vspace*{1pt} model is completed by specifying
inverse-gamma prior $\sigma^2 \sim\IGam(a_{\sigma}, b_{\sigma})$ on
variance parameter $\sigma^2$.

\subsection{Many\hspace*{4pt} existing\hspace*{4pt} structure\hspace*{4pt} alignment\hspace*{4pt} algorithms\hspace*{4pt} are\hspace*{4pt} special\hspace*{4pt} cases}\label{sec3.4}\label{seinter}
Rather~than summarize the posterior $P(M \mid X,Y)$ by Monte Carlo
sampling, we may instead obtain a single MAP alignment (\ref{EqnMAP})
by maximizing the (log-) posterior. Conditioning on parameters $\theta
= (\sigma^2,g,h,\lambda)$, we obtain
%
%
\begin{eqnarray}\label{eqlogpost}
&& \log\bigl(P(M\mid X,Y,\theta)\bigr)\nonumber
\\
&&\qquad = -\frac{3}{2}\llvert M\rrvert\log(2 \pi\sigma) - \frac{1}{2\sigma^2} d_P^2(X_M,Y_M)
+ \bigl(n + m - \llvert M\rrvert\bigr)\log(\lambda)
\\
&&\quad\qquad{} + \log\bigl(Z(g,h)\bigr) - u(M;g,h)\nonumber
\end{eqnarray}
and noting that $\sum_{i=1}^{s(M)} l_i(M) = (n+m) - 2\llvert
M\rrvert$, this is
equivalent to minimizing
\[
d_P^2(X_M,Y_M)
+ u\bigl(M;g^*,h^*\bigr) + C\bigl(\sigma^2,
\lambda, g, h\bigr), 
\]
where $g^{*}=\sigma^2(g + \frac{3}{2}\log(2\pi\sigma) + \log
\lambda)$
and $h^{*}=\sigma^2 h$, and $C(\sigma^2, \lambda, g, h)$ is
independent of $M$. Therefore, the MAP estimate for $M$ with
$(g,h,\lambda,\sigma^2)$ fixed corresponds to a global alignment
obtained via dynamic programming [\citet{NeWu70}], using RMSD under
optimal least-squares rotation/translation as the dissimilarity
metric, and with gap opening and extension penalties given by $g^*$ and
$h^*$. Since $g \ge0$, the term
$(\frac{3}{2}\log(2\pi\sigma) + \log\lambda)$ serves as a
lower bound on $g^{*}$, the ``effective'' gap extension penalty.

Note that the relative posterior probability of two alignments which
differ by an unmatched pair $(x_i,y_j)$ is greater than one if and only if
\[
\bigl\llvert y_j - \bigl(x_i{R} + {\mu}'
\bigr)\bigr\rrvert< g^{*}(1-\xi_{ij}) + h^{*}
\xi_{ij},
\]
where $\xi_{ij}$ is an indicator taking value 1 if removing the pair
$(x_i,y_j)$ creates a new gap in the alignment and 0 otherwise. Thus,
the model favors inclusion of pairs below a dynamically estimated
threshold given by $g^{*}$ and $h^{*}$. Since these threshold
parameters are estimated from the data (in contrast to standard
optimization-based alignment algorithms where they are fixed {a~priori}), our approach automatically controls for the error rates
associated with multiple comparisons.

It is also worth noting that the assumption of normally distributed
errors in (\ref{EqnShapeModel}) may be replaced with an alternative
error model, altering the $d_P$ term in (\ref{EqnStructJoint}) and
the corresponding posterior distribution. In particular, robust error
models with heavy tails may be considered (e.g., Student-$t$ or double
exponential distributions) to account for possible outliers. In this
way, our probabilistic formulation provides statistical insight into
various existing optimization-based algorithms.

For example, \citet{GeLe98} define the similarity between residues
$x_i$ and $y_j$ by
\[
S_{ij} = \frac{c}{1 + ( {d_{ij}}/{d_0} )^2}, %
\]
where $d_{ij}$ denotes the distance between $i$ and $j$ under the
current optimal registration and $c$ and $d_0$ are arbitrarily chosen
constants. Then dynamic programming is employed to obtain the
alignment $M$ maximizing the similarity between proteins, defined by
$\sum_{(i,j)\in M} S_{ij}$. This is equivalent to obtaining the MAP
estimate under our Bayesian model when the distribution of the error
$\varepsilon$ is given by $f(x) \propto\exp\{ -M (1 +
(x/d_0^2) )^{-1} \}$, which is an exponentiated Cauchy
density. [This is indeed a proper density as
$\int_{-\infty}^{\infty} \exp\{ -\frac{M}{ 1 + ({x}/{d_0}
)^2 } \} \,dx < \infty$.] Thus,\vspace*{1pt} our unified probabilistic
framework allows us to interpret such heuristics in terms of their
underlying assumptions about the data generating process.


\section{Computational algorithms}\label{sec4}\label{secomp}
Combining (\ref{EqnStructJoint}), (\ref{eqpriormatrix}) and
(\ref{eqpriorgh}), we obtain the posterior distribution,
\[
P\bigl(M,g,h,\sigma^2 \mid X, Y\bigr) \propto P\bigl(X,Y \mid M,
\sigma^2\bigr)P(M \mid g,h) P\bigl(\sigma^2\bigr) P(g)
P(h).
\]
This posterior can be explored using a Markov chain Monte Carlo
algorithm that iterates between sampling from the conditional
distributions, $P(M \mid g,h, \sigma^2, X, Y)$, $P(g,h \mid M, X, Y)$ and
$P(\sigma^2 \mid M, X, Y)$. The full-conditional posterior for the
variance $\sigma^2$ is obtained by
standard conjugate updating:
\[
\sigma^2 \mid M,X,Y \sim\IGam\bigl( a_{\sigma} +
\tfrac{3}{2}\llvert M\rrvert, b_{\sigma} + \tfrac{1}{2}
d_P^2(X_M,Y_M) 
\bigr).
\]
The gap penalty parameters $(g,h)$ are updated jointly by a
two-dimensional geometric random walk proposal with
Metropolis--Hastings acceptance probability
%
%
\begin{equation}\label{EqnGHAccept}
\qquad 1 \wedge
\frac{Z(g',h') \me^{-u(M;g',h')}}{Z(g,h) \me^{-u(M;g,h)}} \frac
{g'h'}{gh} 
\biggl(\frac{g'}{g} \biggr)^{a_g-1} \biggl(\frac{h'}{h}
\biggr)^{a_h-1} \me^{- (b_g (g' - g)+b_h (h' - h) )}.
\end{equation}
In the previous expression $(g', h')$ correspond to the proposed values
and $(g,h)$ correspond to the current values of the gap parameters. The
required normalizing constants $Z(g,h)$ can be calculated
efficiently via the recursions provided in \hyperref[SecDP]{Appendix}.

As shown by \citet{Schmidler2003}, if we \textit{condition} on
registration parameters $(R,\mu)$, the alignment matrix $M$ may be
sampled from its full conditional distribution using a
forward--backward algorithm similar to that of sequence alignment
[\citet{ZhLiLa98,LiLa99}] and described in \hyperref[SecDP]{Appendix}.
\citet{Wang2006} use this approach for structural alignment.
However, here we have instead defined the likelihood
(\ref{EqnStructJoint}) directly on shape space using maximal values
$(\hat{R}_M,\hat{\mu}_M)$ associated with each distinct $M$, so this
is no
longer the case. But we may still use this efficient block Gibbs step
to generate efficient Metropolis--Hastings proposals
$P(M \rightarrow M')$ with distribution $q(M';R_M,\mu_M)$, where
$(R_M,\mu_M)$ is the registration associated with the
\textit{current} state $M$, and
\begin{eqnarray*}
&& q\bigl(M';R_M,\mu_M\bigr)
\\
&&\qquad \propto P
\bigl(M' \mid g,h\bigr) \bigl(2\pi\sigma^2
\bigr)^{-{3\llvert M'\rrvert}/{2}} \me^{-{1}/{(2\sigma^2)}
\llVert Y_{M'} - \hat{X}_{M',M}\rrVert_F^2} \prod_{y_i \in Y_{\bar{M'}}}
f(y_i \mid\lambda),
\end{eqnarray*}
where $\hat{X}_{M',M} = X_{M'}\hat{R}_M + \mathbf{1}\hat{\mu
}_{M}'$. This $q$
can be sampled efficiently using the recursions of
\hyperref[SecDP]{Appendix}. The proposed $M'$ is then accepted according to
the Metropolis--Hastings criteria
\[
1 \wedge
\frac{P(X,Y \mid M',\sigma^2)P(M' \mid g,h) q(M;R_{M'},\mu
_{M'})}{P(X,Y \mid M,\sigma^2)P(M \mid g,h)q(M';R_M,\mu_M)},
\]
with the required normalizing constants of $q$ obtained from the
sampling recursions.


These dynamic programming proposals are highly efficient for local
sampling and sufficient for closely matched proteins. However, when
multiple alternative alignments with distinct rotation/translations
exist, mixing between them will be slow. We therefore add an
additional Metropolized\vspace*{1pt} independence step where global moves are
proposed without conditioning on the values of $(\hat{\mu}_M,
\hat{R}_M)$ associated with the current alignment. To construct the
independence proposal distribution, we first generate a library of viable
registrations using the following procedure:
\begin{longlist}[2.]
\item[1.] Compute the least-squares registration for each pair of
consecutive 6-residue subsequences on protein $X$ to each such
subsequence on $Y$.
\item[2.] If the subsequence RMSD is less than threshold $\delta$, include
the corresponding registration in the library.
\end{longlist}
This library is computed once upon initialization of the algorithm
and stored for use throughout the simulation, generating an efficient
data-set-specific proposal distribution that deals effectively with
potential multimodality in the posterior. A proposal is made from
this distribution by drawing a registration $(R', \mu')$ uniformly at
random from
the library and proposing a new alignment $M'$ from $q(M';R',\mu')$
using the forward--backward algorithm. It is then accepted according to
the Metropolis--Hastings criteria
\[
1 \wedge
\frac{P(X,Y \mid M',\sigma^2)P(M' \mid g,h) q(M;R',\mu')}{P(X,Y \mid
M,\sigma^2)P(M \mid g,h)q(M';R', \mu')},
\]
%
%
leaving the posterior distribution invariant.

\section{Bayesian synthesis of sequence and structure information}\label{sec5}

Another advantage of the Bayesian probabilistic framework given above
is the ability to seamlessly incorporate additional information when
available. For example, our approach leads to a natural algorithm for
performing alignments based on primary sequence and tertiary structure
simultaneously. This approach allows alignments which synthesize two
types of
information: geometric conservation of the protein architecture, and
physico-chemical properties and evolutionary information provided by
sidechain identities. As an important consequence, our approach
enables the estimation of evolutionary distances from \textit{structure
comparison}, which has previously been quite difficult [\citet
{ChLe86,JoSuBl90,Gr97,LeGe98,WoPe99,KoLe02}]. Being able to estimate
evolutionary distances from
structural information has important implications because structure
is much more strongly conserved than sequence, enabling comparisons
across much longer evolutionary timescales.

The model given by (\ref{EqnStructJoint}) for structural observations
is easily extended to account simultaneously for both sequence \textit{and} structure information by assuming the structure and sequence to
be conditionally independent given the alignment $M$, that is,
$P(A^x,A^y,X,Y\mid M,\theta) = P(A^x,A^y\mid M,\theta)P(X,Y\mid
M,\theta)$. We take the conditional likelihood of the sequences given
the alignment to be
%
%
\begin{equation}
P\bigl(A^x,A^y \mid M,\Theta\bigr) = \prod
_{(i,j) \in M}\Theta\bigl(A^x_i,A^y_j
\bigr) \prod_{i \notin M }\Theta\bigl(A^x_i,
\cdot\bigr) \prod_{j \notin M }\Theta\bigl(
\cdot,A^y_j\bigr), \label{EqnSeqJoint}
\end{equation}
where $A_i^y$ is the $i$th amino acid in protein $x$, $\Theta(a,b)$
gives the probability of residues $a$ and $b$ being matched on related
sequences, and $\Theta(a,\cdot) = \Theta(\cdot,a)$ gives the marginal
probability for residue $a$. Equation (\ref{EqnSeqJoint}) is the standard
likelihood form of sequence alignment [\citet{Thompson86}], and these
joint and marginal distributions form the bases of standard sequence
alignment substitution matrices such as PAM and BLOSUM
[\citet{Dayhoff68,Henikoff92}], where the distributions are estimated
from alignments of closely related proteins. For example, the PAM-$k$
substitution can be written as $\Psi_{k}=(\Psi_k(a,b))$, where
\[
\Psi_k(a,b) =10\log_{10} \biggl(\frac{\Theta_k(a,b)}{\Theta(a,\cdot
)\Theta
(\cdot,b)}
\biggr),
\]
and $k$ represents the expected percentage of amino acid
replacements, most often between 30 and 250, with larger numbers
used for sequences further away in the evolutionary scale. Sequence
alignment may then be formulated as a maximum-likelihood or Bayesian
inference problem [\citet{Durbin1998,ZhLiLa98,LiLa99,Thompson86}],
where the introduction of $\Theta_k$ amounts to the introduction of
a number of new fixed hyperparameters. Inference on $k$ can also be
carried out by placing a prior distribution over members of the PAM
family of matrices [\citet{ZhLiLa98}]. The posterior distribution on
$k$ then provides an estimate of evolutionary distance between the
two proteins.

Multiplication of equations (\ref{EqnStructJoint}) and
(\ref{EqnSeqJoint}) directly yields a joint likelihood for inferring
$M$ by combining both sequence and structure information. However, as
we noted in Section~\ref{sec2}, structure is generally much more strongly
conserved than
sequence, thus, we would like to weight the contribution of structure
information in determining the alignment more heavily than that of
sequence. In this way the sequences will serve primarily to provide
supplementary information in regions of the alignment where
structural information leaves uncertainty; as we will see, it also
permits the estimation of evolutionary distance from the largely
structure-based alignment.

To control the relative weighting of sequence and structure
information, we introduce a concentration (or inverse temperature) parameter
$\eta$, resulting in the modified sequence likelihood
\begin{eqnarray*}
&& \Pr\bigl(A^x,A^y \mid M,\Theta,\eta\bigr)
\\
&&\qquad  =
\frac{\prod_{(i,j) \in M }
\Theta(A^x_i,A^y_j)^{\eta}
\prod_{i \notin M }\Theta(A^x_i,\cdot)^{\eta} \prod_{j \notin M }
\Theta(\cdot,A^y_j)^{\eta}}{
\sum_{A^{x*},A^{y*}} \prod_{(i,j) \in M }
\Theta(A^{x*}_i,A^{y*}_j)^{\eta}
\prod_{i \notin M }\Theta(A^x_i,\cdot)^{\eta} \prod_{j \notin M }
\Theta(\cdot,A^y_j)^{\eta}}
\\
&&\qquad  = \prod_{(i,j) \in M} \frac{\Theta(A^x_i,A^y_j)^{\eta}}{\sum
_{A_r,A_s}\Theta
(A_r,A_s)^{\eta}} \prod
_{i \notin M} \frac{\Theta(A^x_i,\cdot)^{\eta}}{\sum_{A_r}
\Theta(A_r,\cdot)^{\eta}}\prod_{j \notin M}
\frac{\Theta(\cdot,A^y_j,)^{\eta}}{\sum_{A_s}
\Theta(\cdot,A_r)^{\eta}}.
\end{eqnarray*}
Setting $\eta=1$ corresponds to simple multiplication of the
sequence and structure likelihoods (\ref{EqnStructJoint}) and
(\ref{EqnSeqJoint}), while as $\eta\rightarrow0$, $\Pr(A^x,A^y
\mid M,\Theta,\eta)$ approaches a uniform distribution for every
$\Theta$ and $\eta=0$ reduces to the structure-only model
(\ref{EqnSeqJoint}). Thus, $\eta^{-1}$ plays the role of
a dispersion parameter for the discrete observations $A$.
We can consider estimating $\eta$ directly under
the restriction $\hat{\eta} < 1$, in which case $\hat{\eta}$ can
be interpreted as a measure of agreement between sequence and
structure, which shrinks to down-weight the contribution of sequence
information if sequence and structural information are in conflict.

\section{Examples}\label{sec6}\label{seexamples}
We apply our Bayesian structural alignment algorithm to a number of
illustrative examples. Hyperparameter\vspace*{1pt} values used are given in
Table~\ref{tahyperpar}: the prior distribution for $\sigma^2$ has mean
1.5~$\mbox{\AA}$ and variance 1.0~$\mbox{\AA}$, in line with the results
for analogous proteins in \citet{ChLe86}, and following \citet{GeLe98},
the prior mean for $h$ is about 40 times larger than the prior mean for
$g$. Results were mostly unaffected by changes in the prior mean for
$\sigma^2$ between 0.5~$\mbox{\AA}$ and 4.0~$\mbox{\AA}$, or by changes in the prior
mean of $g$ and $h$ of around 50\%. All
inferences described are based on 100,000 samples obtained after a
burn-in period of 20,000 iterations, with convergence verified by
visual inspection of the trace plots and using the Gelman--Rubin
convergence test [\citet{GeRu92}]. Monitored quantities include the
length of the alignment, the rotation angles corresponding to rotation
matrix $\hat{R}_{M}$, the translation vector $\hat{\mu}_{M}$ and the
two gap penalty parameters $(g,h)$. We report MAP alignments unless
otherwise noted.
%
%
\begin{table}
\tabcolsep=0pt
\tablewidth=200pt
\caption{Hyperparameter values used in the examples}\label{tahyperpar}
\begin{tabular*}{\tablewidth}{@{\extracolsep{\fill}}@{}lccccc@{}}
\hline
$\bolds{a_{\sigma}}$ & $\bolds{b_{\sigma}}$ & $\bolds{a_h}$ & $\bolds{b_h}$ & $\bolds{a_g}$ & $\bolds{b_g}$\\
\hline
2.25 & 1.5 & 2 & $1/2$ & 2 & 20 \\
\hline
\end{tabular*}
\end{table}

We first analyze 16 pairs of proteins from \citet{OrStOl02}. This list
includes pairs of very different lengths and proteins from various
structural classes, including $\alpha$ proteins containing primarily
$\alpha$-helical secondary structure, $\beta$~proteins containing primarily
$\beta$-sheets and $\alpha+ \beta$ proteins containing significant
fractions of
both. Table~\ref{tacomparison} summarizes the results
obtained using three different values for $\lambda$ ranging from a
relatively low (7.6) to the relatively high (9.6), and compares the
Bayesian alignments against those obtained using the popular CE
algorithm [\citet{ShBo98}]. In most cases, the differences between
Bayesian and CE alignments are important; in more than half the
examples less than 20\% of the matched residues coincide. These
differences are mostly due to the way CE handles gaps: to reduce the
computational complexity, CE assumes that gaps cannot be introduced
simultaneously in both proteins. Similar restrictions can be easily
introduced in our model [by setting $q_{i,j}(2,3) = 0$ in \hyperref[SecDP]{Appendix}], and when this
is done the results for both methods tend to agree. Generally
speaking, the added flexibility means that the quality of the Bayesian
alignments is superior to CE: it tends to produce alignments
containing more matched residues but with a lower RMSD.
\begin{sidewaystable}
\tabcolsep=0pt
 \tablewidth=\textwidth
\caption{Bayesian structural alignments of the 16 pairs of proteins
from \citet{OrStOl02}, compared against the popular CE algorithm
[\citet{ShBo98}]. $\llvert M\rrvert$ is the length of the alignment,
$N_{\mathrm{CE}}$ denotes the number of matches in common with
CE, and RMSD is expressed in~$\mbox{\AA}$. The values of $g$ and $h$ reported
correspond to posterior means}\label{tacomparison}
\begin{tabular*}{\tablewidth}{@{\extracolsep{\fill}}@{}ld{3.0}d{3.0}d{3.0}cccd{2.0}ccd{3.0}cd{2.0}ccd{3.0}cd{2.0}cc@{}}
\hline
& & & \multicolumn{2}{c}{\textbf{CE}} & \multicolumn{5}{c}{$\bolds{\lambda=7.6}$}& \multicolumn{5}{c}{$\bolds{\lambda=8.6}$}& \multicolumn{5}{c}{$\bolds{\lambda=9.6}$}
\\[-6pt]
& & & \multicolumn{2}{c}{\hrulefill} & \multicolumn{5}{c}{\hrulefill}& \multicolumn{5}{c}{\hrulefill}& \multicolumn{5}{c}{\hrulefill}
\\
\textbf{$\bolds{X}$--$\bolds{Y}$} & \multicolumn{1}{c}{$\bolds{n}$} &
\multicolumn{1}{c}{$\bolds{m}$} & \multicolumn{1}{c}{$\bolds{|M|}$} &
\multicolumn{1}{c}{\textbf{RMSD}} & \multicolumn{1}{c}{$\bolds{|M|}$} &
\multicolumn{1}{c}{\textbf{RMSD}} & \multicolumn{1}{c}{$\bolds{N_{\mathbf{CE}}}$} &
\multicolumn{1}{c}{$\bolds{h}$} & \multicolumn{1}{c}{$\bolds{g}$} &
\multicolumn{1}{c}{$\bolds{|M|}$} & \multicolumn{1}{c}{\textbf{RMSD}} &
\multicolumn{1}{c}{$\bolds{N_{\mathbf{CE}}}$} & \multicolumn{1}{c}{$\bolds{h}$} &
\multicolumn{1}{c}{$\bolds{g}$} & \multicolumn{1}{c}{$\bolds{|M|}$} &
\multicolumn{1}{c}{\textbf{RMSD}} & \multicolumn{1}{c}{$\bolds{N_{\mathbf{CE}}}$} &
\multicolumn{1}{c}{$\bolds{h}$} & $\bolds{g}$\\
\hline
1ABA--1DSB & 87 & 188 & 56 & 4.5 & 24 & 2.2 & 0 & 9.64 & 0.01 & 57 & 3.7 & 0 & 6.26 & 0.13 & 76 & 4.7 & 14 & 6.10 & 0.42\\
1ABA--1TRS & 87 & 105 & 70 & 2.7 & 65 & 3.0 & 37 & 5.68 & 0.14 & 72 & 3.4 & 38 & 5.68 & 0.22 & 75 & 3.6 & 38 & 5.70 & 0.24\\
1ACX--1COB & 108 & 151 & 92 & 4.0 & 66 & 2.1 & 49 & 6.89 & 0.06 & 86 & 3.8 & 57 & 6.60 & 0.15 & 93 & 4.1 & 57& 6.38 & 0.21 \\
1ACX--1RBE & 108 & 104 & 56 & 7.3 & 25 & 2.5 & 6 & 9.02 & 0.01 & 31 & 2.8 & 0 & 7.89 & 0.02 & 50 & 4.2 & 15 & 7.45 & 0.03 \\
1MJC--5TSS & 69 & 194 & 61 & 2.7 & 52 & 2.3 & 25 & 7.89 & 0.03 & 60 & 3.0 & 29 & 7.24 & 0.36 & 66 & 3.9 & 15 & 7.36 & 0.44\\
1PGB--5TSS & 56 & 194 & 48 & 2.9 & 39 & 2.3 & 19 & 6.52 & 0.56 & 55 & 3.3 & 40 & 6.60 & 0.87 & 55 & 3.1& 34& 6.65 & 0.94\\
1PLC--1ACX & 102 & 108 & 80 &3.3 & 71 & 3.4 & 23 & 5.92 & 0.10 & 84 & 4.0 & 23 & 5.80 & 0.20 & 89 & 4.6 & 23 & 6.30 & 0.22 \\
1PTS--1MUP & 119 & 157 & 80 & 4.1 & 76 & 3.0 & 0 & 6.80 & 0.06 & 83 & 3.1 & 0 & 6.60 & 0.09 & 88 & 3.5 & 0 & 6.77 & 0.12 \\
1TNF--1BMV & 152 & 185 & 115 & 4.1 & 70 & 2.7 & 3 & 7.86 & 0.02 & 109 & 4.2 & 40 & 7.10 & 0.08 & 113 & 4.3 & 35& 6.94 & 0.11 \\
1UBQ--1FRD & 76 & 98 & 64 & 4.4 & 62 & 3.0 & 28 & 5.41 & 0.15 & 62 & 2.9 & 23 & 5.10 & 0.25 & 65 & 3.1 & 32 & 5.36 & 0.29 \\
1UBQ--4FXC & 76 & 98 & 64 & 4.0 & 46 & 2.3 & 22 & 5.46 & 0.13 & 61 & 2.9 & 34 & 5.20 & 0.24 & 66 & 3.4 & 42 & 5.43 & 0.29 \\
2GB1--1UBQ & 56 & 76 & 48 & 3.1 & 44 & 2.1 & 0 & 5.38 & 0.18 & 51 & 3.4 & 6 & 5.27 & 0.46 & 51 & 3.3 & 6 & 5.66 & 0.59 \\
2GB1--4FXC & 56 & 98 & 48 & 3.6 & 35 & 3.5 & 0 & 9.06 & 0.06 & 53 & 3.9 & 7 & 7.56 & 0.42 & 55 & 4.1 & 0 & 7.55 & 0.62 \\
2RSL--3CHY & 119 & 128 & 80 & 4.1 & 43 & 2.6 & 22 & 7.99 & 0.02 & 76 & 3.8 & 31 & 6.76 & 0.08 & 81 & 4.0 & 33 & 6.67 & 0.11\\
2TMV--256B & 154 &106 & 84 & 3.5 & 65 & 2.3 & 0 & 7.39 & 0.06 & 79 & 2.9 & 0 & 7.13 & 0.10 & 89 & 4.0 & 69 & 6.87 & 0.19\\
3CHY--1RCF & 128 & 169 &116 & 3.9 & 80 &3.0 & 38 & 6.89 & 0.06 & 122 & 4.5 & 87 & 5.99 & 0.54 & 126 & 4.7 & 70 & 6.05 & 0.76 \\
\hline
\end{tabular*}
\end{sidewaystable}

Some pairs of proteins seem to be somewhat sensitive to the choice of
$\lambda$ (e.g., 1ACX--1COB), while the alignment of other pairs
seems to be remarkably robust (e.g., 1UBQ--FRD). In general,
larger values of $\lambda$ (which imply larger penalties for both
opening and extension) tend to generate longer alignments. The
sensitivity of the model to $\lambda$ is not surprising; indeed,
setting $\lambda= 0$ immediately implies that the optimal alignment is
empty for any pair of proteins. The model is robust to small changes in
$\lambda$ (around 5\% or so), which can be absorbed by $g$ and $h$.
However, when $\lambda$ is increased by a large amount, we set a new
baseline threshold that pairs of residues need to satisfy in order to
be included in the alignment, favoring the alignment of sections that
were previously not close enough to be aligned. When structural
similarity varies dramatically along the alignment (as is the case for
some pairs in our list), this ``threshold effect'' can produce
important changes in the resulting alignment. In general, we can think
of $\lambda$ as controlling the tightness of the alignment. In our
experience, a conservative value such as $\lambda= 7.6$ works well in
most applications, and we use this value in further illustrations.

Table~\ref{tacomparison} also shows the posterior median of the gap
penalties for each alignment. Opening penalties range between 5 and 9,
while extension penalties range from 0.01 to nearly 0.9, reflecting the
differing levels of similarity across different pairs.
%
%

\begin{figure}

\includegraphics{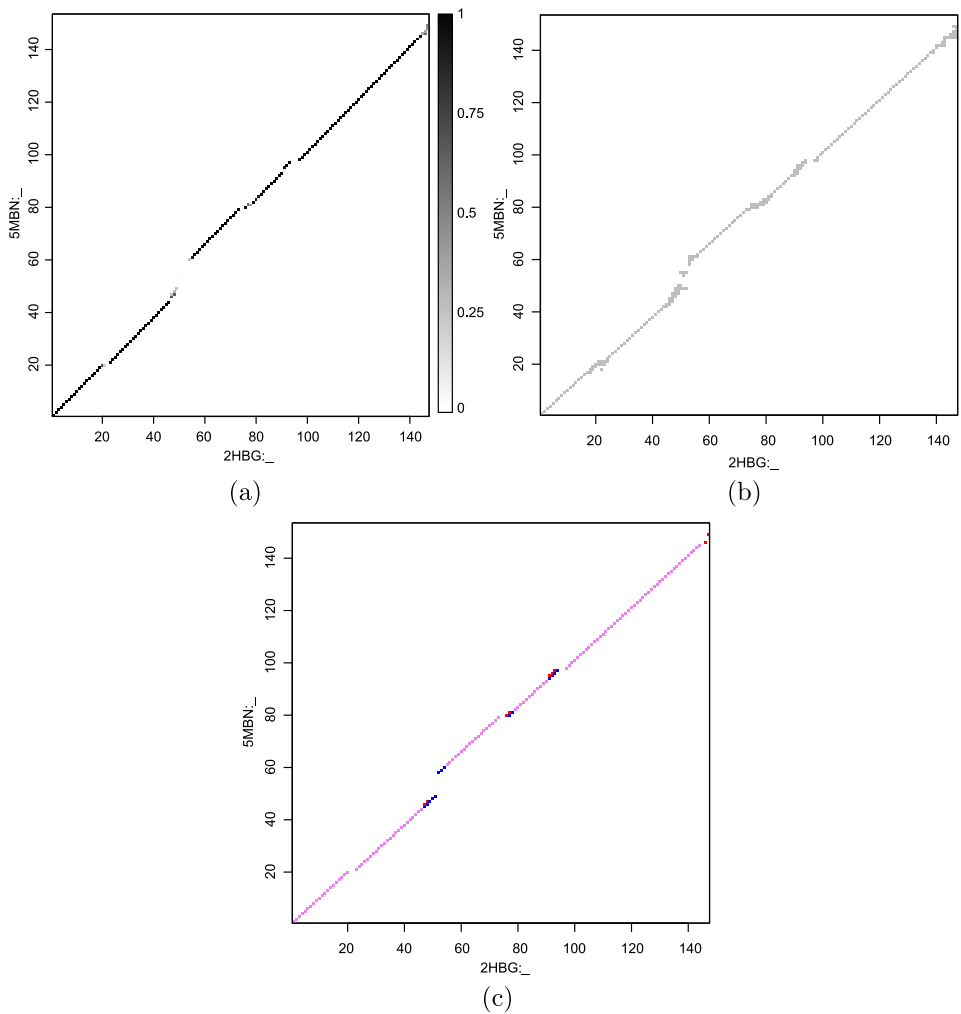}

\caption{Bayesian structural alignment of 5MBN and 2HBG.
\textup{(a)}~Marginal alignment probability matrix for all pairs of
residues, showing uncertainty associated with the alignment.
\textup{(b)}~Plot of all sampled alignments.
\textup{(c)}~Comparison of the MAP alignment (red) with the CE alignment (blue);
common regions are shown in purple.}\label{fi5mbn2hbg}
\end{figure}

\begin{figure}

\includegraphics{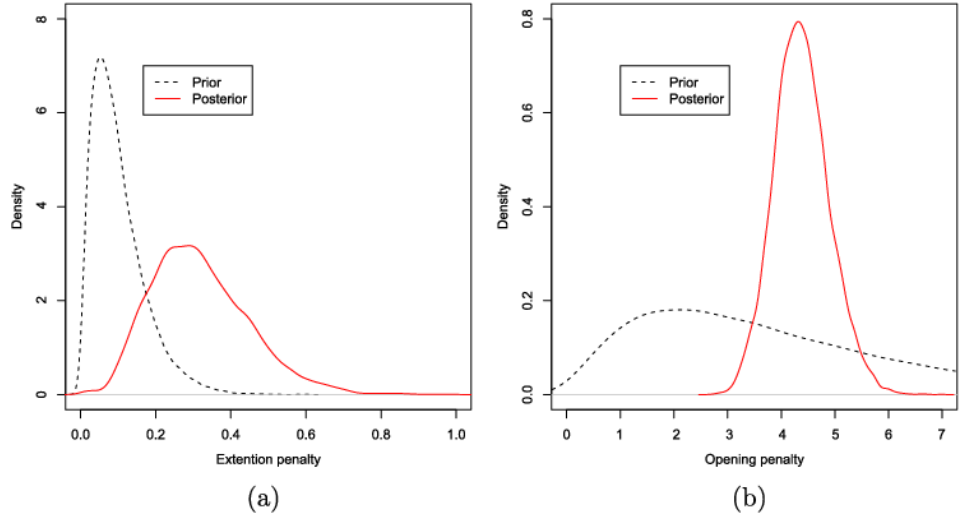}

\caption{Prior and posterior distributions for gap
penalty parameters obtained for the Bayesian alignment of globins
5MBN and 2HBG. The Bayesian approach allows the algorithm to
adaptively determine the appropriate gap parameters rather than
treating them as fixed.}\label{figappenalties}
\end{figure}

Next, we consider in detail the alignment of two $\alpha$ proteins
from the globin family, 5MBN and 2HBG. Figure~\ref{fi5mbn2hbg}
presents both the marginal alignment matrix (which provides
information on the uncertainty associated with the alignment) and the
MAP alignment, comparing it against that obtained from CE. The most
striking feature about this example is that different alignment
methods tend to disagree in regions where the uncertainty in the
Bayesian alignment is high (the regions surrounding the gap between
residues 47 and 62 of 5MBN, the extremes of the helix between
residues 81 and 98 of 5MBN, and at the very end of the alignment).
This highlights the importance of using a probabilistic alignment
framework, rather than relying on a single
optimum. Figure~\ref{figappenalties} shows the prior and posterior
distributions for both gap penalty parameters in this
example, which demonstrate that the model does adaptively estimate
parameters relevant to the data at hand.

%
%
%
%
%

%
\begin{figure}

\includegraphics{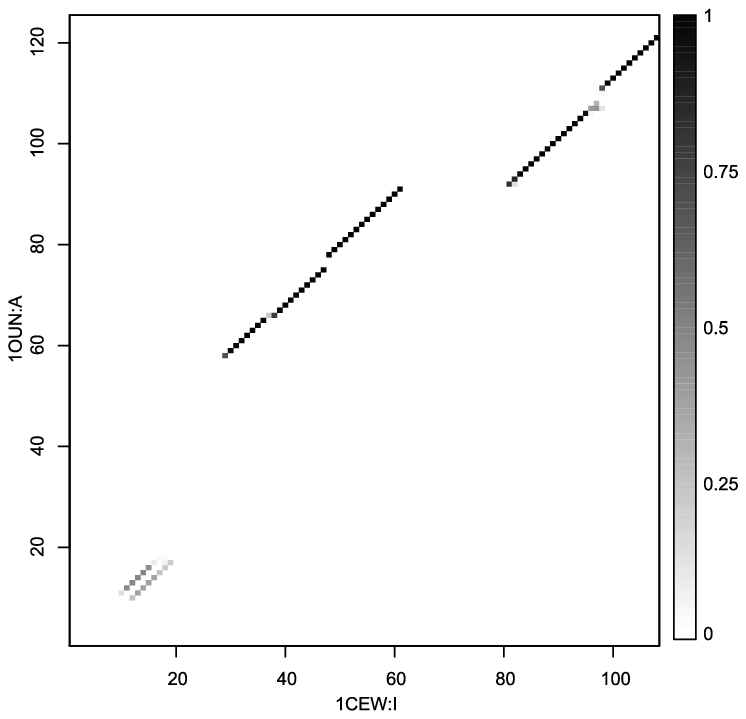}

\caption{Marginal alignment matrix for the Bayesian structural
alignment of 1OUN:A and 1CEW:I. The posterior uncertainty in the
alignment can be seen at the N-terminus, where two possible
alignments of the $\alpha$-helix at positions 10--20 have comparable
posterior probabilities.}\label{fiouncewmar}
\end{figure}

Finally, we explore the alignment of the $\alpha+ \beta$ proteins
1CEW\_I and 1OUN\_A. \citet{LaKoSiDo00} describe two alternative
alignments for these proteins having a comparable number of equivalent
residues (70 vs. 68) and RMSD (2.4~$\mbox{\AA}$  both), which arise by shifts
in the alignment of the secondary structures. Figure~\ref{fiouncewmar} shows the marginal alignment probabilities for all
pairs of residues. Unlike the previous example, uncertainty levels in
this alignment are very high, particularly in the $\alpha$-helix
region between residues 10 and 20. The two alternative alignments for
this region correspond to the two alignments described in
\citet{LaKoSiDo00}. However, the alignment of the rest of the proteins
corresponds to the 70 residue alignment discussed by those authors. This
example shows how the global sampling of the full posterior enables
the model to automatically weight the relative importance of closely
related alternative alignments, and how the estimation of gap
penalties can further improve this.
%

%

\subsection{Combined sequence--structure alignment}\label{sec6.1}\label{SecResults}
To illustrate the performance of our simultaneous
sequence-and-structure alignment approach, we consider two pairs of
proteins that have been previously analyzed in the literature. For
convenience we consider a discrete set of discount factors ranging
from 0 to 1 in increments of 0.1, along with 21 PAM matrices ranging
from PAM100 to PAM300. ``Noninformative'' uniform prior distributions
are used for both discount factors and PAM matrices. All results are
based on 130,000 iterations of the Gibbs sampler, after a burn-in
period of 30,000 iterations.

\begin{figure}

\includegraphics{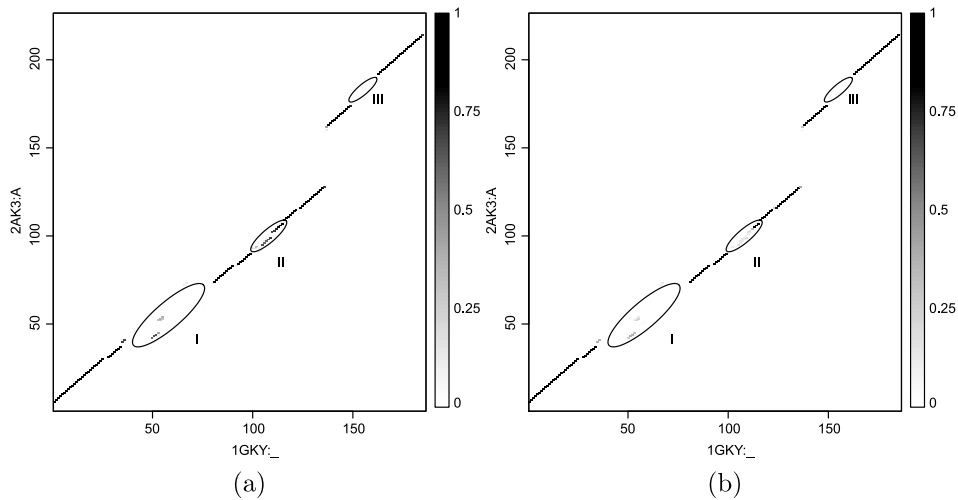}

\caption{Marginal probabilities over aligned pairs for 1GKY and
2AK3\_A. \textup{(a)} Shows alignments based only on structure, while
\textup{(b)}~presents alignments that also incorporate sequence information.
Although there is some structural similarity in regions \textup{I}~and~\textup{II},
sequence similarity in these areas is low (see Table~\protect\ref{alignsecII}).}\label{gyak}
\end{figure}

In the first example we analyze two kinases studied by Bayesian
sequence alignment in \citet{ZhLiLa98}; a guanylate kinase from yeast
(1GKY) and an adenylane kinase from the beef heart mitochondrial
matrix (2AK3\_A), which are VAST structural neighbors [\citet{vast96}].
A structural alignment for these proteins using the combinatorial
extension (CE) algorithm [\citet{ShBo98}] shows very little sequence
similarity (under 13\% identity). Figure~\ref{gyak} compares our
structural and simultaneous sequence--structure alignments for these
two proteins by showing the marginal probability of aligning any pair
of residues integrated over all other parameters in the model
(including PAM matrices and discount factors). Both algorithms tend
to agree on which regions should be aligned. For example, both avoid
aligning the section of the $\alpha$-helix located between residues
150--162 in 1GKY and residues 175--191 in 2AK3\_A (marked III in
Figure~\ref{gyak}). The axes for these helixes are not parallel,
producing a
big divergence at the C terminus.
%
%
%

%
%
\begin{table}
\tablewidth=300pt
\tabcolsep=10pt
\caption{Sequence alignment of corresponding to the best structural
alignment between region \textup{II} of 2AK3\_A and 1GKY, with residues 93--100
of 2AK3\_A matched with residues 103--111 of 1GKY. Numbers correspond
to the PAM 250 (log-odds) scores for each matched residue pair and
clearly show that despite the shape similarity, there is little
evidence of common ancestry in this region of the protein}\label{alignsecII}
\begin{tabular}{l|cccccccc}
2AK3\_A & R & T & L & P & Q & A & E & A \\
1GKY & G & V & K & S & V & K & A & I \\ \hline
& $-$3 & 0 & $-$3 & 1 & $-$2 & $-$1 & 0 & $-$1
\end{tabular}
\end{table}

In spite of the similarities, some differences are evident among both
models. For example, a section of the alignment starting at residue
108 of 1GKY (marked II in Figure~\ref{gyak}) is excluded when the
sequence information is included in the analysis. Both proteins
present a short helix in this region, and they can be structurally
aligned reasonably well. However, there are important
incompatibilities in the two sequences for these helices, which
suggests that this section is not functionally important. Table~\ref
{alignsecII} presents the sequence correspondence associated with
the structural alignment of this section, along with the scores for
each site. Note that the structural alignment implies no conserved
residues in the area and the substitution of various basic and acidic
amino acids by either hydrophobic or hydrophilic residues. Indeed, of
the eight substitutions, only one happens between members of a common
chemical group. This is a local discrepancy between sequence and
structure that is not seen in other regions of the proteins, and
suggests that the region should be dropped from the alignment. Similarly,
a couple of short regions in the remote site for mono and triphosphate
binding located between residues 35--80 for 1GKY and 38--73 in 2AK3\_A
(marked I in Figure~\ref{gyak}), that show a moderate probability of
being aligned under the structural model, are down-weighted (but not
completely removed) when the sequence information is included. This
region, which was probably functionally important in an ancestor, has
degraded since both proteins diverged and does not seem functionally
active in these proteins. These two minimal changes in the alignment
lowers the RMSD from 3.5~$\mbox{\AA}$ to a median of 1.95~$\mbox{\AA}$ [with a 90\% high
posterior density interval of (1.84, 2.17)].

%
%
\begin{figure}

\includegraphics{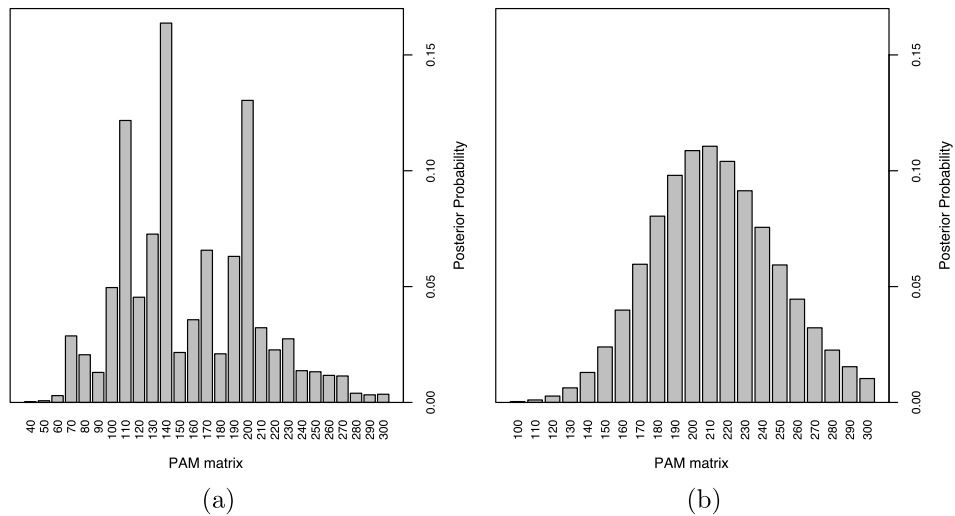}

\caption{Posterior probabilities of PAM distances based on sequence
information alone \textup{(a)} and based on the Bayesian sequence--structure
alignment.}\label{pPAM}
\end{figure}
%

Figure~\ref{pPAM} shows the marginal posterior probability
distribution over PAM matrices that arises from our joint
sequence--structure model, contrasting
it with the results in \citet{ZhLiLa98}. Whereas the sequence-based
analysis in the original paper led to a multimodal posterior with
modes at PAMs 110, 140 and 200, our posterior is smooth and unimodal,
with its mode located between PAM200 and PAM210. This demonstrates
the strong additional information obtained by aligning based on
structure and sequence simultaneously: virtually all sequence
alignments which are compatible with structural alignment indicate
the larger evolutionary distance (posterior mean 212, median
206). The marginal mode for the temperature is 1 (posterior
probability 0.57), indicating that there is very little need to
discount sequence with respect to structure information.

%
%
\begin{figure}[b]

\includegraphics{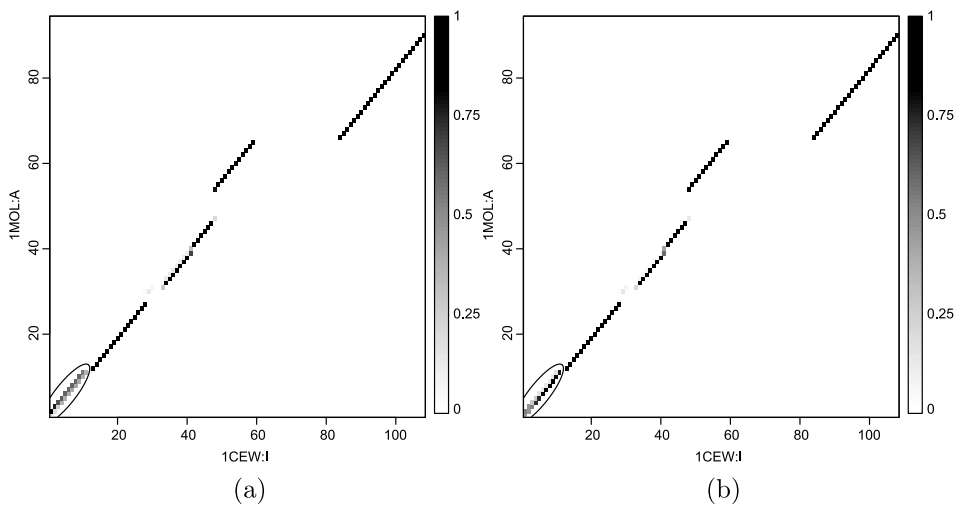}

\caption{Marginal probabilities over aligned pairs for 1MOL\_A and
1CEW\_I. \textup{(a)} Shows alignments based only on structure, while \textup{(b)}
presents alignments that also incorporate sequence information.
Circles show the strand region discussed in Table~\protect\ref{molcewcomp}.}
\label{molcew}
\end{figure}

Our second example focuses on comparing the single-chain fused
Monellin from the Serendipity berry (1MOL\_A) and the chicken egg
white Cystatin (1CEW\_I) analyzed previously in \citet{LaKoSiDo00}
and \citet{KoNiTe03}. Figure~\ref{molcew} shows the Bayesian
alignments obtained with and without inclusion of sequence
information. Again, the two alignments are quite similar as expected,
but the sequence information leads to small refinements in the
structural alignment. For example, two alternative alignments of the
initial strand are supported by the structure-only alignment, with the one
where 1MOL\_A is shifted toward the C terminus being slightly
preferred (this is also the one preferred by CE). However,
incorporation of sequence information reverses this to prefer the
N-terminus shifted alignment (approximate posterior probabilities of
0.85 vs 0.15),
and examination of the sequences strongly supports this choice.
Table~\ref{molcewcomp} shows the sequence alignment under both
alternatives, with amino acids colored by a simple classification
according to physico-chemical properties (Table~\ref{aaclas})
to demonstrate the improved similarity on top of
amino acid identity.
The sequence--structure alignment yields six matches in amino acid
type, including an additional two identities and a hydrophilic match
on top of the three hydrophobics achieved by the structural alignment.
The corresponding price paid in structural distance [mean RMSD of
1.91~$\mbox{\AA}$ versus 1.89~$\mbox{\AA}$, with both 90\% h.p.d. regions being (1.81~$\mbox{\AA}$, 2.05~$\mbox{\AA}$)]
is insignificant. This example clearly shows that incorporation of
sequence information can refine structural alignments in areas where
the structure alignment is ambiguous.

%
%
\begin{table}
\tablewidth=247pt
\caption{Sequence alignment of the first strand of 1MOL\_A and 1CEW\_I
induced by the two alternative models. \textup{(a)} Mode using structural
information only and \textup{(b)} mode under the Bayesian simultaneous
sequence--structure alignment. Colors refer to the classification in
Table~\protect\ref{aaclas}; note the improvement in matching of chemical
classes}\label{molcewcomp}

\includegraphics{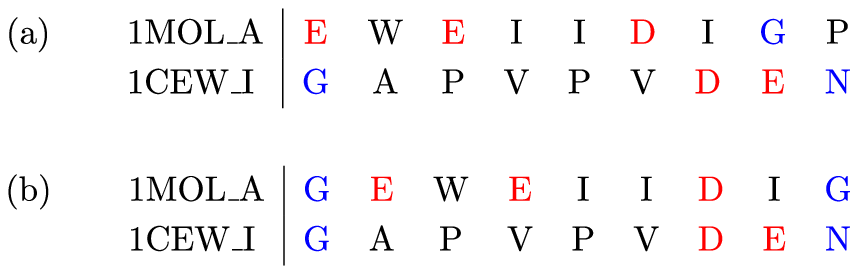}

\end{table}

%
%
\begin{table}[b]
\caption{Simple amino acid classification based on chemical
properties}\label{aaclas}

\includegraphics{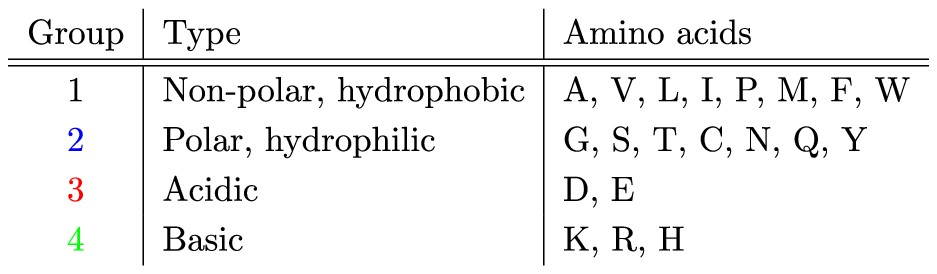}

\end{table}

\begin{figure}

\includegraphics{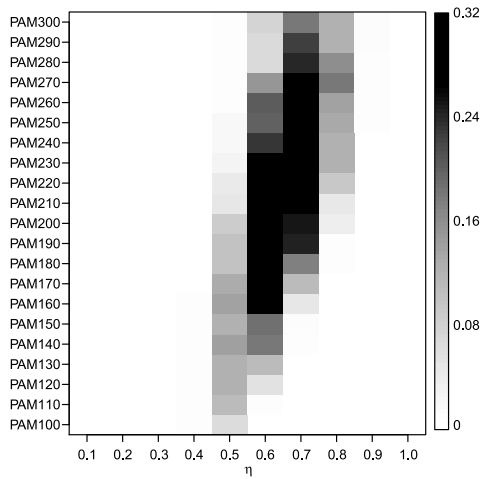}

\caption{Heat map representation of the joint posterior distribution over
discount factors and PAM matrices for 1MOL\_A and
1CEW\_I.}\label{jpplambdaPAMmol-cew}
\end{figure}

Figure~\ref{jpplambdaPAMmol-cew} shows the joint posterior
distribution over PAM matrices and discount factors for this example.
Relative to the previous example, there is more uncertainty in both
the evolutionary distance and the discount factor. The diagonal
pattern in the plot suggests an obvious dependence between these
two parameters. This is to be expected, as both $\eta$ and
evolutionary distance increase the entropy of the joint amino acid
distribution. Nevertheless, the results point toward a relatively
large divergence time (recall one is a plant protein and the other
is an animal protein), with the mode of the distance at 210.
%
%
%

To avoid confounding of PAM and tempering parameters, one parameter
may be chosen in advance and fixed. For example, the substitution
matrix may be chosen to reflect prior information about evolutionary
distance and inference performed only on the \textit{discount factor} or
vice-versa. When 1MOL\_A and
1CEW\_I are aligned using PAM250 as the fixed substitution matrix, the
resulting distribution for \textit{discount factor} is very similar: the
mode is located at $\eta=0.6$ with a posterior probability of 0.32,
and most of the remaining mass concentrates in $\eta=0.5$ and $\eta=0.7$,
both with posterior probability of 0.24. Differences in the actual
alignments are not obvious from the marginal distribution plot (not
shown). However, a more detailed look at the values shows that fixing
the PAM matrix further decreases the probability of the CE-like
alignment below 10\%.

\begin{figure}

\includegraphics{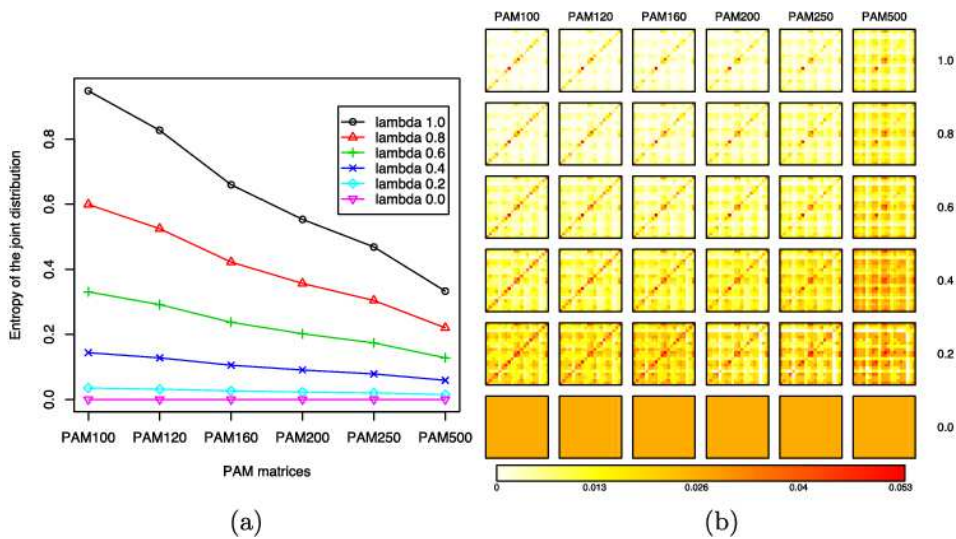}

\caption{\textup{(a)} Entropies of the joint distribution induced by different
evolutionary distances and tempering parameters. \textup{(b)} Heat map plots
of the joint distributions. Amino acids are ordered alphabetically,
starting with Alanine in the lower left.}
\label{FigPAMentropies}
\end{figure}

The examples discussed in this section show that simultaneous
estimation of PAM distance and discount factor may be difficult. Since
larger evolutionary distances increase sequence divergence/decrease
sequence conservation, both low discount values and high PAM distances
imply more tolerance to substitutions. One way to measure substitution
tolerance is via the Shannon entropy of the joint distributions
(Figure~\ref{FigPAMentropies}). Although increasing $\eta$ and $k$
both increase this entropy, they do so in slightly different ways. We
observe that PAM100 with a temperature of 0.8 has roughly equivalent
entropy to untempered PAM200. However, PAM100/0.8 assign a much larger
probability of match than does PAM200/1.0, as seen by the darker
diagonal. In addition, temperature increases treat all combinations of
amino acids in the same way, and thus low probability regions tend to
disappear quickly. This is not so for increases in the evolutionary
distances. In the limit of $k$, the PAM joint distribution will
converge to the product of independent marginal distributions given by
the stationary distribution of the underlying Markov chain (estimated
as overall population frequencies). In contrast, as the discount factor
approaches 0, the joint distribution and thus the marginal
distributions converge to uniform. Figure~\ref{FigPAMentropies}(a) also
shows that differences between PAM matrices grow weaker as the discount
factor decreases. In the extreme case when $\eta=0$, all matrices are
equivalent.
%
%
%

Finally, it is important to mention that we have not found alternative
methodologies in the literature capable of this type of information
synthesis, against which to compare our results. One of the few
methods available is an extension of the combinatorial extension (CE)
method \citet{ShBo98}, accessible via
\surl{http://cl.sdsc.edu/ce.html}. However, in this implementation there is
little control on the choice of substitution matrices and, for the
examples we have studied, the sequences seems to have little practical
influence in the final results.

\section{Conclusions}\label{sec7}
We have presented a unifying probabilistic framework for protein
structure alignment based on Bayesian hierarchical modeling.
Computationally efficient MCMC algorithms for sampling the posterior
distribution enable us to directly account for uncertainty over
alignments, including identification of alternative alignments and
evaluation of their relative importance. Our model provides insights
into the relations between and assumptions of standard
optimization-based alignment techniques, along with a unifying
framework that facilitates comparisons between them. It also naturally
incorporates additional information, such as the inclusion of sequence
information in structural alignments. As a byproduct of the latter,
we obtained a model which can estimate evolutionary distance directly
from structural alignment, an otherwise difficult task. The examples
shown clearly highlight how these advantages of our model aid in
identification of functionally relevant regions and in resolving
ambiguities in alignments. By introducing a discount parameter, we
are able to control the influence of the sequence information on the
final alignment, an important characteristic missing in previous
attempts to combine sequence and
structure. As noted, PAM distance and discount factor are correlated,
and inference on evolutionary distance will therefore be more reliable
if additional information is used to determine the discount factor;
this is an area for additional study. Finally, we feel that
sequence--structure alignments provide the most insight when used
in conjunction with structure-only alignments as done in the examples.
Comparisons between the two appear to provide more direct information
on conservation than do comparisons between structure-only and
sequence-only alignments.

\begin{appendix}
\section*{Appendix: Dynamic programming forward--backward sampling}\label{sec8}
\label{SecDP}
As shown by \citet{Schmidler2003}, if we \textit{condition} on
registration parameters $(R,\mu)$, the alignment matrix $M$ may be
sampled from its full conditional distribution using a
forward--backward algorithm similar to that of sequence alignment
[\citet{ZhLiLa98,LiLa99}]. Let $v_{i,j}(k)$ be the probability of the
alignment of the $i$th prefix of $X$ and the $j$th prefix of $Y$
ending in type $k$, with $k=1$ meaning that both final residues are
aligned, $k=2$ inserts a gap in $X$ and $k=3$ inserts a gap in $Y$.
Then
\begin{eqnarray*}
v_{i,j}(1) & =& \sum_{k=1}^{3}
q_{i,j}(k,1)v_{i-1,j-1}(k),\qquad v_{i,j}(2)  = \sum
_{k=1}^{3} q_{i,j}(k,2)v_{i-1,j}(k),
\\
v_{i,j}(3) & =& \sum_{k=1}^{3}
q_{i,j}(k,3)v_{i,j-1}(k) 
\end{eqnarray*}
and letting $d_{ij}^2 =\llVert y_j - (x_i{R} + \mathbf{1}{\mu
}')\rrVert^2$, the
transition weights are given by
%
\[
q_{i,j}(l,k) = \cases{
\displaystyle\frac{\lambda}{(2 \pi\sigma
^2)^{{3}/{2}}} \exp\biggl\{ -\frac{1}{2\sigma^2}d_{ij}^2 \biggr\}, &\quad$k=1$,
\vspace*{5pt}\cr
\displaystyle\exp\{ g+h \}, &\quad$(l,k) = (1,2)$ or $(1,3)$,
\vspace*{3pt}\cr
\displaystyle\exp
\{ g \}, &\quad$(l,k) = (2,2)$ or $(3,3)$ or $(2,3)$,
\vspace*{3pt}\cr
0, &\quad$(l,k) =
(3,2)$.}
\]
In order to ensure identifiability of the alignments, we do not allow
a gap in $Y$ to follow a gap in $X$, hence, $q_{3,2}=0$. The
initialization of these recursions are
%
\begin{eqnarray*}
v_{1,1}(1) & =& \frac{\lambda} {(2 \pi\sigma^2)^{3/2}} \exp\biggl\{
-\frac{1}{2\sigma^2}
d_{11}^2 \biggr\},
\\
v_{i,1}(1) & =& \frac{\lambda} {(2 \pi\sigma^2)^{3/2}} 
\exp\biggl\{ -
\frac{1}{2\sigma^2} d_{1i}^2 + (i-1)g + h \biggr\},
\\
v_{1,j}(1) & =& \frac{\lambda} {(2 \pi\sigma^2)^{3/2}} \exp\biggl\{
-\frac{1}{2\sigma^2}d_{j1}^2
+ (j-1)g + h \biggr\},
\\
v_{1,j}(2) & =& \exp\bigl\{ (j+1)g + h \bigr\}\quad\mbox{and}\quad
v_{i,1}(3) = 0.
\end{eqnarray*}
Note that $v_{n,m}$ contains the sum over all alignments and, given
$(g,h)$, the same algorithm with $q_{i,j}(l,1) = 1$ can be used to
efficiently compute the normalizing
constant $Z(g,h)$ in the gap-penalty prior (\ref{eqpriormatrix}), as
required for the acceptance probability (\ref{EqnGHAccept}). Once
$q_{i,j}(k)$ is available for all $(i,j)$, the alignment is sampled
backward, starting with
\[
u_{n,m}(k) = \frac{v_{n,m}(k)}{\sum_{l=1}^{3} v_{n,m}(l)}
\]
and then conditionally adding a matched pair or a gap on one of
the proteins with probabilities:
\[
u_{i,j}(k,l) = \frac{q_{i-1,j-1}(l,k)v_{i-1,j-1}(k)}{\sum
_{k=1}^{3}q_{i-1,j-1}(l,k)v_{i-1,j-1}(k)}.
\]
\end{appendix}

\section*{Acknowledgments}
The authors would like to thank the Editors and one anonymous referee
for numerous suggestions that greatly improved the quality of the manuscript.


%

\printaddresses
\end{document}